\documentstyle[prl,aps,epsf,multicol]{revtex}

\begin{document}

\newcommand{\centreline}{\centerline}

\draft  

\title{The Superconducting Gap in {MgB$_{2}$}: \\
       Electronic Raman Scattering Measurements of Single Crystals}

\author{J.\ W.\ Quilty, S.\ Lee, A.\ Yamamoto and S.\ Tajima}
\address{Superconductivity Research Laboratory, 
        International Superconductivity Technology Center,\\ 
        1-10-13 Shinonome, Koto-ku, Tokyo 135-0062, Japan}

\maketitle

\begin{abstract}

Polarisation-resolved Raman scattering measurements were performed on
MgB$_{2}$ single crystals to determine the magnitude, symmetry and
temperature dependence of the superconducting gap. A single sharp peak
due to Cooper pair breaking appears in the electronic continuum below
$T_c$, reaching an average maximum Raman shift of $105 \pm
1$~cm$^{-1}$ ($2\Delta(0) / k_{B}T_{c} = 3.96 \pm 0.09$) and showing
up to 5~cm$^{-1}$ anisotropy between polarised and depolarised
spectra. The temperature dependence of $2\Delta$ follows that
predicted from BCS theory, while the anisotropy decreases with
decreasing temperature. It is concluded that the Raman results are
consistent with a slightly anisotropic \textit{s}-wave gap in a
conventional BCS superconductor.

\end{abstract} 

\pacs{PACS numbers: 74.25.Gz,
                    74.70.Ad,
                    78.30.Er}

\begin{multicols}{2}
\narrowtext

The recently discovered superconductor MgB$_{2}$
\cite{Nagamatsu01:nat401} shows an unusually high $T_c$ for a simple
binary compound, immediately raising questions regarding the nature of
superconductivity in this material
\cite{An01:prl86,Kortus01:prl86,Liu01:prl87,Yildirim01:prl87,%
Imada01:jpsj70,Furukawa01:jpsj70}. Experiments on polycrystalline
samples generally support a gap of \textit{s}-like symmetry, but
measurements of the gap magnitude vary widely. Some groups report a
single superconducting gap, with gap-to-$T_{c}$ ratios ranging from
0.6 to 5
\cite{Takahashi01:prl86,Rubio-Bollinger01:prl86,Sharoni01:prb63,%
Kotegawa01:prl87}, while other groups find multiple gap-like features
with gap-to-$T_{c}$ ratios around 1.1--1.8 and 3.5--4.8
\cite{Szabo01:prl87,Giubileo01:prl87,Tsuda01:prl87,Chen01:prl87}. Due
to the difficulties and uncertainties associated with polycrystal
measurements, these results strongly motivate further study of the gap
feature(s) in single crystals.

Electronic Raman scattering is a useful experimental technique for
studying superconductors because it provides a direct probe of the
superconducting gap via the breaking of Cooper pairs by the incident
light \cite{Klein84:prb29}. The frequency, polarisation dependence and
temperature dependence of the pair breaking peak reflect the
magnitude, symmetry and temperature dependence of the gap
\cite{Klein84:prb29,Devereaux95:prb51:22,Devereaux95:prl74}. Knowledge
of these fundamental properties is essential for the underlying
mechanism of superconductivity to be deduced. For polycrystalline
samples, the MgB$_{2}$ Raman spectrum presents a broad phonon feature
superimposed on a relatively intense electronic continuum
\cite{Chen01:prl87,Goncharov01:prb64,Hlinka01:prb64}, the latter
undergoing a superconductivity-induced renormalization which has been
interpreted in terms of two gaps of magnitudes ($2\Delta$)
44~cm$^{-1}$ and 100~cm$^{-1}$ \cite{Chen01:prl87}. In this letter we
report the first Raman measurements of the fundamental gap properties
in single crystals of MgB$_{2}$.

Single crystals of MgB$_{2}$ were synthesized from a precursor mixture
of 99.9\% pure magnesium powder and 97\% pure amorphous boron. The
crystals were grown in a BN container under 4--6~GPa pressure at
1400--1700$^{\circ}$C for 5 to 60 minutes. Magnetization and
resistivity measurements give $T_{c} = 38.2$~K with a transition width
of 0.3~K \cite{Lee01:jpsj70}. Four circle XRD measurements indicate
these are high quality single crystals, with no evidence of bulk
impurities. Slab-like crystals with relatively clean and flat in-plane
surfaces were selected for measurement. These crystals exhibited
broken hexagonal shape of typical dimension 200--300~$\mu$m square by
30--40~$\mu$m thick. The crystals were mounted for in-plane
measurement and oriented with the \textit{a}-axis horizontal, within
$\pm 5^{\circ}$, confirmed by x-ray diffraction measurements. Further
confirmation was provided by the results of superconducting-state
Raman measurements. The incident light was polarised either vertically
(V) or horizontally (H), that is, either perpendicular or parallel to
the \textit{a}-axis. Only vertically polarised scattered light was
collected. In the $D_{6h}$ point group, VV polarisation selects
excitations of $A_{1g}$ and $E_{2g}$ symmetry, while HV polarisation
selects only the $E_{2g}$ component. Henceforth the VV and HV
configurations will be referred to as \emph{polarised} and
\emph{depolarised}.

Raman spectra were measured with a Jobin-Yvon T64000 triple
monochromator, charge coupled device, and an excitation wavelength of
514.5~nm provided by an Ar-Kr laser. The typical incident power was
0.8~mW, point-focussed to a spot of 100~$\mu$m diameter. To determine
the degree of laser overheating, spectra were also measured at laser
powers higher than typical. The only noticeable effect of higher
incident powers was spot heating, and where appropriate subsequently
quoted temperatures have been corrected for this effect. Temperature
control was provided by a closed-cycle helium refrigerator and at
typical laser powers sample overheating was estimated to be less than
2~K. All spectra have been corrected for the Bose thermal
contribution.

\begin{figure}
    \centreline{\epsffile{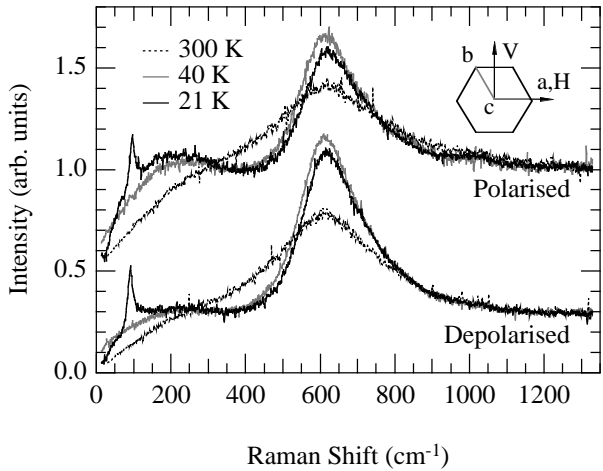}}
    \vspace{0.25cm}
    \caption{Polarised (VV) and depolarised (HV) Raman spectra of
    crystal~A at 300, 40 and 21~K. Inset: polarisation orientation 
    relative to the {MgB$_{2}$} crystallographic axes.}
    \label{fig:Overview}
\end{figure}

Polarised and depolarised Raman spectra of a single crystal of
MgB$_{2}$ (crystal~A) from 15 to 1300~cm$^{-1}$ are shown in
Fig.~\ref{fig:Overview}. The spectral intensity has been normalised at
high wavenumbers and the polarised spectra
\noindent
offset by 0.5 units along the \textit{y}-axis. Dominating the spectra
is a very broad, asymmetric feature located at around 620~cm$^{-1}$,
which is commonly attributed to the Raman-active E$_{2g}$ symmetry
phonon in MgB$_{2}$ \cite{Goncharov01:prb64,Hlinka01:prb64}. This
feature narrows substantially with decreasing temperature, consistent
with suggestions that the E$_{2g}$ phonon is very anharmonic
\cite{Yildirim01:prl87}, but curiously shows no indication of a large
superconductivity-induced hardening below $T_c$ predicted by band
theoretical calculations \cite{Liu01:prl87}. Cursory modelling of this
feature with a Fano equation, an unsatisfactory description of the
observed spectral lineshape, revealed 7--10~cm$^{-1}$ hardening
between 40 and 21~K. Further experiments must be performed to clarify
the character of this mode.

In contrast, a clear superconductivity-induced renormalization of the
electronic continuum is visible in the 21~K spectra, where a sharp
pair breaking peak forms at around 100~cm$^{-1}$. The magnitude of the
renormalization is greater in depolarised spectra. Accompanying the
peak is a decrease of scattering intensity at lower Raman shifts,
although some scattering intensity remains directly below the peak. In
the case of an isotropic \textit{s}-wave gap the theoretically
calculated electronic Raman spectrum is characterised by a pair
breaking peak located at an energy of $2\Delta$ in all scattering
channels, below which there is zero scattering intensity
\cite{Klein84:prb29}. For anisotropic \textit{s}-wave the scattering
threshold remains, located at the minimum gap energy, while pair
breaking peaks appear at different Raman shifts in different
polarisations, reflecting the gap anisotropy
\cite{Devereaux95:prb51:22,Devereaux95:prl74}. Since the polarisation
dependence of the peak seen in MgB$_{2}$ is not strongly anisotropic,
we directly associate the pair breaking peak energy with $2\Delta$.

\begin{figure}
    \centreline{\epsffile{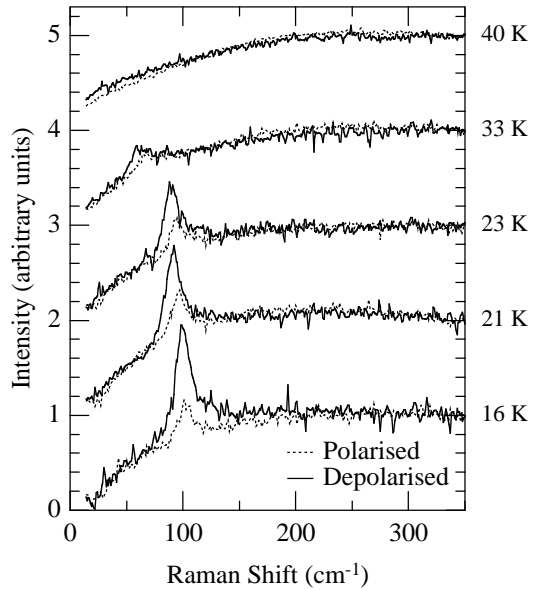}}
    \vspace{0.25cm}
    \caption{Polarised and depolarised Raman spectra of 
    crystal~A at temperatures $T \leq T_{c}$. The pair breaking peak
    sharpens and rises in energy with decreasing temperature.}
    \label{fig:CrystalA}
\end{figure}

Figure~\ref{fig:CrystalA} shows Raman spectra from crystal~A in the
vicinity of the pair breaking peak. The spectra have been normalized
to unity around 350~cm$^{-1}$ and successive spectra have been
progressively offset by one unit along
\noindent
the \textit{y}-axis. Where appropriate, temperatures have been
corrected for sample heating effects. Here, the evolution of the
$2\Delta$ peak with temperature is seen and a small anisotropy in peak
energy between polarised and depolarised spectra is apparent. In the
vicinity of 25~cm$^{-1}$ the Raman intensity at low temperatures
nearly vanishes. To obtain an estimate of the gap energy $2\Delta(T)$,
the spectrum in the vicinity of the pair breaking feature was
approximated by a Gaussian with a cubic background, and fitted with a
least squares algorithm. Subsequently quoted uncertainties in
$2\Delta(T)$ are the 95\% confidence intervals obtained from the fit.
The average anisotropy between the two polarisations thus obtained is
$4 \pm 1$~cm$^{-1}$, the polarised $2\Delta$ peak always appearing
higher in energy than the depolarised peak. Close examination of
Fig.~\ref{fig:CrystalA} suggests that the anisotropy is about
5~cm$^{-1}$ at temperatures close to $T_c$, and falls with decreasing
temperature to a little less than 3~cm$^{-1}$ at the lowest
temperatures measured.

\begin{figure}
    \centreline{\epsffile{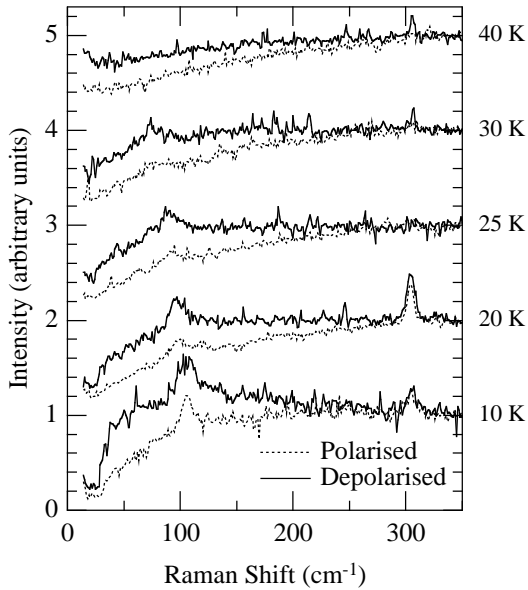}}
    \vspace{0.25cm}
    \caption{Polarised and depolarised Raman spectra of 
    crystal~B at temperatures $T \leq T_{c}$. Here, an additional
    feature may be seen below the pair breaking peak.}
    \label{fig:CrystalB}
\end{figure}

A number of single crystals were measured and low-frequency Raman
spectra from a second crystal, crystal~B, are shown in
Fig.~\ref{fig:CrystalB}. In common with crystal~A, a pair breaking
peak appears in the superconducting state spectrum and $2\Delta(T)$
agrees well between crystals. Anisotropy in $2\Delta$ is also present,
amounting to $2 \pm 1$~cm$^{-1}$ on average, about half that seen in
crystal~A. Some additional structure is also apparent. A sharp peak
present in most crystal~B spectra near 300~cm$^{-1}$, but not present
in crystal~A, is attributable to surface contamination. Although the
polarised spectra are comparable between crystals, additional
scattering intensity is present in the depolarised spectra of
crystal~B. At low temperatures a
\noindent
sharp shoulder forms around 30--40~cm$^{-1}$, which is very similar to
that seen in polycrystal results \cite{Chen01:prl87}. Residual
scattering intensity in the vicinity of 25~cm$^{-1}$ is non-vanishing
in crystal~B and the magnitude of the superconductivity-induced
renormalization in the depolarised spectra is reduced compared to
crystal~A. Inspection of crystal~B with an optical microscope showed a
less regular surface and confirmed the presence of surface impurities.
Typically, crystals with flat, clean surfaces showed a sharp $2\Delta$
peak with low residual scattering while crystals with evidence of
surface disorder showed higher residual scattering and a shoulder at
30--40~cm$^{-1}$.

Within the usual theoretical description of non-resonant electronic
Raman scattering, the square of the Raman vertex $\gamma_{\mathbf
k}^{2}$ weights scattering from different portions of the Fermi
surface depending on the incident and scattered light polarisations
\cite{Devereaux95:prb51:22}. Taking the band dispersion given by the
nearest neighbour tight binding model \cite{Kortus01:prl86}
we made a straightforward calculation of $\gamma_{\mathbf k}^{2}$ for
$A_{1g}$ and $E_{2g}$ symmetries and found that the former weights
excitations from almost the entire hexagonal Brillouin zone,
particularly near the vertices, while the latter weights excitations
around the edges. Thus, the observation of a smaller gap magnitude in
depolarised ($E_{2g}$) spectra suggests an anisotropic in-plane gap of
the form $\Delta_{\mathbf k} = \Delta_{0} [1 + \epsilon
\cos(6\theta)]$, compatible with \textit{s}-wave symmetry
\cite{Seneor02:prb65}.

In determining the superconducting gap symmetry, and thus placing
constraints on the underlying mechanism of superconductivity, the
temperature dependence of the gap provides another convenient test.
BCS theory gives an explicit temperature dependence for a
\textit{s}-wave gap with which the experimental data may be compared.
\begin{figure}
    \centreline{\epsffile{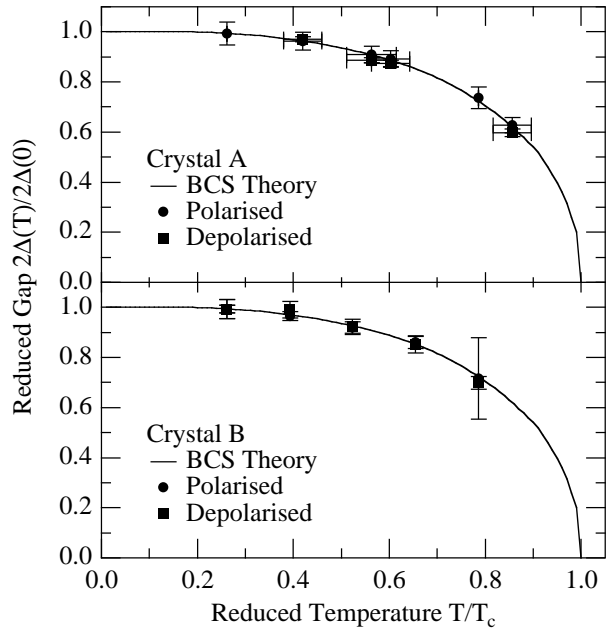}}
    \vspace{0.25cm}
    \caption{Reduced gap temperature dependence for crystals A and B.
    The solid line is the reduced gap from BCS theory.}
    \label{fig:ReducedGap}
\end{figure}
\noindent
We calculate the reduced gap $2\Delta(T) / 2\Delta(0)$ from our data,
extrapolating $2\Delta(0)$ from our lowest temperature measurements.
Results of this calculation are shown in Table~\ref{tab:MaxGap}.
\begin{table}
    \begin{tabular}{llll}
                  &              & $2\Delta(0)$&  \\
        Sample    & Polarisation & (cm$^{-1}$) & $2\Delta(0)/k_{B}T_{c}$ \\ \hline
        Crystal A & Polarised    & $106 \pm 2$ & $3.99 \pm 0.09$  \\
                  & Depolarised  & $103 \pm 1$ & $3.88 \pm 0.05$  \\
        Crystal B & Polarised    & $106 \pm 1$ & $4.01 \pm 0.05$  \\
                  & Depolarised  & $105 \pm 1$ & $3.96 \pm 0.08$  \\
    \end{tabular}
    \caption{Extrapolated values of maximum gap energy and the 
             corresponding gap to $T_c$ ratio.}
    \label{tab:MaxGap}
\end{table}
\noindent
The magnitude of the superconducting gap energy, averaged between
polarisations, is $2\Delta(0) = 3.96 \pm 0.09$, which is consistent
with MgB$_{2}$ being a moderately strong coupling superconductor.
Figure~\ref{fig:ReducedGap} shows $2\Delta(T) / 2\Delta(0)$ for both
crystals, which follow the BCS-predicted temperature dependence.
Uncertainties in temperature are shown for those data points to which
correction was applied. The temperature dependence of the
superconducting gap, as directly measured by Raman scattering,
indicates that MgB$_{2}$ is a conventional BCS superconductor.

The presence of Raman scattering intensity below 2$\Delta$ seen in
Figs.~\ref{fig:CrystalA} and \ref{fig:CrystalB} initially appears to
contravene a simple \textit{s}-wave interpretation of the data. The
theory of Raman scattering in superconductors predicts a zero
intensity threshold below the pair breaking peak in an \textit{s}-wave
superconductor
\cite{Klein84:prb29,Devereaux95:prb51:22,Devereaux95:prl74} and Raman
studies of the Nb$_{3}$Sn and V$_{3}$Si conventional superconductors
\cite{Dierker83:prl50,Hackl83:jpc16,Hackl89:pc162} confirm a sharp
decrease in scattering intensity below the 2$\Delta$ peak. In
contrast, the high-$T_c$ superconductors exhibit broad, anisotropic
pair breaking features which are accompanied by significant in-gap
scattering intensity in all scattering channels, a consequence of the
nodes in the \textit{d}-wave gap function
\cite{Devereaux95:prb51:22,Devereaux95:prl74}. Nodes in the gap
function of MgB$_{2}$ might explain the in-gap intensity seen below
T$_{c}$ in crystal~A (Fig.~\ref{fig:CrystalA}), however this
possibility is discounted by the near-isotropic polarisation
dependence of the narrow pair breaking peak. While polycrystal
measurements show gap-like features below 100~cm$^{-1}$
\cite{Chen01:prl87}, association of the 40~cm$^{-1}$ shoulder with a
second superconducting gap is problematic since it is not seen
consistently in both crystals.

A number of extrinsic factors appear to contribute to the in-gap
scattering present in Figs.~\ref{fig:CrystalA} and \ref{fig:CrystalB}.
Rayleigh scattering, impurity and luminescence components, due to
surface contamination, were present to varying degrees during the
measurements, and the low-frequency tail of the very broad E$_{2g}$
phonon may also furnish scattering intensity at low frequencies.
Significant surface disorder may introduce additional scattering
contributions and the similar 40~cm$^{-1}$ shoulder seen in
polycrystal \cite{Chen01:prl87} and crystal~B spectra may be related
to a \textit{c}-axis component of the superconducting gap. Such
surface effects may also contribute scattering intensity to the
below-gap Raman spectrum of crystal~A, although to a much lesser
degree. Clearly the 40~cm$^{-1}$ shoulder is not an intrinsic feature
of the in-plane Raman spectrum, related to a second pair breaking
peak. Rather it appears to be an extrinsic effect related to sample
surface quality.

In summary, we have measured the temperature and polarisation
dependence of the $2\Delta$ pair breaking peak seen in the in-plane
electronic Raman continuum of MgB$_{2}$ single crystals. This peak
shows BCS-like temperature dependence with only slight anisotropy,
arguing strongly for conventional \textit{s}-wave superconductivity.
The average gap to $T_c$ ratio $2\Delta / k_{B}T_{c} = 3.96 \pm 0.09$
indicates moderately strong electron phonon coupling. We are able to
identify only one gap feature in the in-plane Raman spectra, and
attribute structure below the pair breaking peak seen in the poorer
crystal to surface defects.

This work was supported by the New Energy and Industrial Technology
Development Organization (NEDO) as collaborative research and
development of fundamental technologies for superconductivity
applications.

\end{multicols}

\end{document}